\documentclass[11pt,preprint2]{aastex}

\usepackage{epsfig}
\usepackage[latin1]{inputenc}

\newcommand{\etal}{et al~}

\shorttitle{Primary Proton Spectrum of Cosmic Rays measured with Single
Hadrons}
\shortauthors{Antoni et al.}

\begin{document}


\title{The Primary Proton Spectrum 
of Cosmic Rays
measured with Single Hadrons at Ground Level
\vspace{0.7cm}
}

\author{
%\small%
T.~Antoni$^1$,
W.D.~Apel$^2$,
F.~Badea$^{2,a}$,
K.~Bekk$^2$,
A.~Bercuci$^{2,a}$,
H.~Bl\"umer$^{1,2}$,
H.~Bozdog$^2$,
I.M.~Brancus$^3$,
C.~B\"uttner$^1$,
A.~Chilingarian$^4$,
K.~Daumiller$^1$,
P.~Doll$^2$,
R.~Engel$^2$,
J.~Engler$^2$,
F.~Fe{\ss}ler$^2$,
H.J.~Gils$^2$,
R.~Glasstetter$^{1,b}$,
A.~Haungs$^2$,
D.~Heck$^2$,
J.R.~H\"orandel$^{1}$,
K-H.~Kampert$^{1,2,b}$,
H.O.~Klages$^2$,
G.~Maier$^2$,
H.J.~Mathes$^2$,
H.J.~Mayer$^2$,
J.~Milke$^2$,
M.~M\"uller$^{2,d}$,
R.~Obenland$^2$,
J.~Oehlschl\"ager$^2$,
S.~Ostapchenko$^{1,c}$,
M.~Petcu$^3$,
H.~Rebel$^2$,
A.~Risse$^5$,
M.~Risse$^2$,
M.~Roth$^1$,
G.~Schatz$^2$,
H.~Schieler$^2$,
J.~Scholz$^2$,
T.~Thouw$^2$,
H.~Ulrich$^2$,
J.~van~Buren$^2$,
A.~Vardanyan$^4$,
A.~Weindl$^2$,
J.~Wochele$^2$,
J.~Zabierowski$^5$
\vspace{0.5cm}
} 

\affil{(1) Institut f\"ur Exp. Kernphysik, Universit\"at
                          Karlsruhe, 76021~Karlsruhe, Germany}
\affil{(2) Institut f\"ur Kernphysik, Forschungszentrum Karlsruhe, 
              76021~Karlsruhe, Germany}
\affil{(3) National Institute of Physics and Nuclear Engineering,
              7690~Bucharest, Romania}
\affil{(4) Cosmic Ray Division, Yerevan Physics Institute,
              Yerevan~36, Armenia}

\affil{(5) Soltan Institute for Nuclear Studies,
              90950~Lodz, Poland \vspace{0.33cm}}
\affil{$^a$ on leave of absence from (3)}
\affil{$^b$ now at: Universit\"at Wuppertal, 42119 Wuppertal, Germany}
\affil{$^c$ on leave of absence from Moscow State University, 
                             119899 Moscow, Russia}
\affil{$^d$ corresponding author, email: mueller@ik.fzk.de}



\begin{abstract}
The flux of cosmic-ray induced single hadrons near sea level has been measured
with the large hadron calorimeter of the KASCADE experiment. The measurement
corroborates former results obtained with detectors of smaller size if the
enlarged veto of the 304 m$^2$ calorimeter surface is encounted for. The
program CORSIKA/QGSJET is used to compute the cosmic-ray flux above the
atmosphere.  Between E$_{0}$~=~300~GeV and 1~PeV the primary proton spectrum
can be described with a power law parametrized as
$dJ/dE_{0}~=~(0.15\pm0.03)~\cdot E_{0}^{-2.78\pm0.03}~
$m$^{-2}$~s$^{-1}$~sr$^{-1}$~TeV$^{-1}$.  In the TeV region the proton flux
compares well with the results from recent measurements of direct experiments.
\end{abstract}




\section{Introduction}
The ``knee'' in the cosmic-ray energy spectrum has been observed by many
research groups and in several observables of air shower experiments.
Typically, its position is found around a primary particle energy of 4~PeV.
Many theoretical approaches to explain the knee exist.  The most probable cause
seems to be a superposition of spectra of many nuclei each with an individual
flux cut-off at different energies. The hypotheses of the origin and
experimental findings, however, differ significantly, in particular for the
primary proton spectrum, which is of special relevance to understand cosmic-ray
acceleration and propagation in the galaxy. A proton knee is claimed to be seen
at different energies, e.g. at 10~TeV by the MUBEE collaboration (Zatsepin
\etal 1993), at 100~TeV by the Tibet group (Amenomori \etal 2000) and at 4~PeV
by the KASCADE collaboration (Ulrich \etal 2001). When taking the experimental
data of older direct measurements by balloon or satellite experiments above the
atmosphere at face value, one might imagine a change of the power law slope at
10~TeV. But recent measurements seem not to confirm such former conjectures.
Precise measurements during the last years in the 100~GeV region yielded proton
fluxes lower by about 30\% compared to older measurements, e.g. the new data of
the BESS (Sanuki \etal 2000), CAPRICE (Mocchiutti \etal 2001), IMAX (Menn \etal
2000) and AMS (Alcaraz \etal 2000) collaborations. On the other hand, in the
100~TeV region recent publications on direct measurements report on higher flux
values, e.g. JACEE (Asakimori \etal 1998) and Runjob (Apanasenko \etal 2001).
These findings indicate that the proton flux does not seem to decrease as
strongly as anticipated so far.

Therefore, it is of interest to determine the proton flux over a wide range of
primary energy using one single method.  Such a method is the detection of
single hadrons at ground level. These unaccompanied hadrons turn out to be
intimately connected to primary protons. The latter penetrate deeper into the
atmosphere with their hadronic component than heavier primaries of the same
energy and are the most abundant producers of single hadrons.  Single hadron
spectra have been measured using different experimental techniques like
emulsion chambers, magnetic spectrometers, or calorimeters. In the past,
measurements have been carried out both, at sea level (Cowan and Matthews 1971,
Siohan et al 1977, Fickle and Lamb 1979, Mielke \etal 1994) and at mountain
altitudes (Aglietta \etal 2003, Inoue \etal 1997).  Different definitions of
single hadrons are used in the literature. For the present investigations a
large calorimeter is used at sea level and single hadrons are defined as
follows: Only one hadron with an energy of at least 100~GeV is reconstructed in
the detector.  In addition, the zenith angles are restricted to less than
30$^\circ$ in the analysis. During the last years the simulation of air showers
has improved considerably and the primary particle spectra can be deduced from
ground based experiments with more confidence.

The KASCADE calorimeter has been operating continuously and steadily for many
years. Large data sets have been accumulated which allow to estimate the flux
up to the PeV range.  However, as air shower simulations indicate, more and
more single hadrons turn out to originate from higher-mass primaries with
increasing energy.  Hence, the connections to primary protons become less
stringent.

\section{Experimental apparatus}

\begin{figure}[t]
\epsfig{file=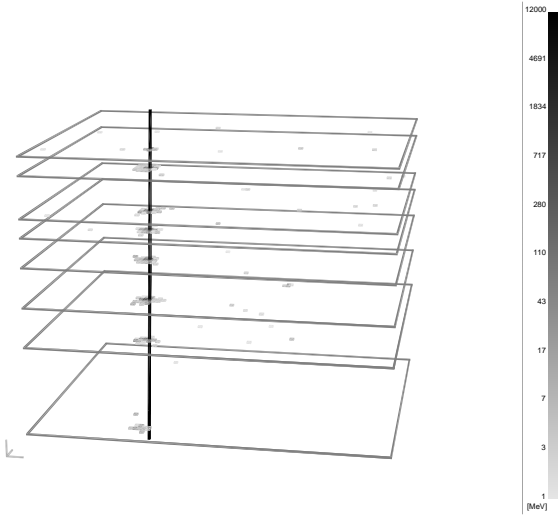,width=0.9\columnwidth,angle=270}
\caption{Pattern of a single hadron event in the calorimeter.
Each pixel represents the deposited energy on an electrode of 25$\times$25~cm$^2$.
The straight line indicates the reconstructed shower axis.}
\label{fig1}
\end{figure}

The hadrons have been detected with the central calorimeter of the KASCADE
experiment measuring cosmic rays near sea level (Engler \etal 1999). It is a
sampling calorimeter consisting of layers of lead, iron, and concrete absorbers
interspersed with 9 layers of warm-liquid ionization chambers with an
acceptance area of 304~m$^2$.  The finely segmented chambers allow to measure
the energy of individual hadrons and to reconstruct their point and angle of
incidence. An example of a single hadron event is presented in figure
\ref{fig1}. The energy depositions in the ionization chambers are plotted from
which a total energy of 4.5~TeV for the hadron shown has been reconstructed.
Apart from the hadron cascade no significant energy deposition is seen in the
calorimeter layers. In particular, also the uppermost layer, where the entire
electromagnetic energy is measured, is nearly empty. The detection and
reconstruction efficiencies of hadronic cascades in the calorimeter have been
determined by simulating cascades with the detector simulation code GEANT (CERN
1993)/FLUKA (Aarnio \etal 1987, 1990).  At 100~GeV a trigger and reconstruction
efficiency of 50\% is reached, increasing to more than 70\% at 500~GeV.
Radiating muons can imitate a hadron. However, above 100~GeV the contributions
of hadrons faked by muons is below the 1\% level, for a detailed discussion see
Mielke \etal (1994).  The maximum energy of an unaccompanied hadron detected
was 50~TeV.  During five years of operation (1996-2001) more than
1.5$\cdot$10$^7$ events have been recorded with at least one reconstructed
hadron, out of which 2.5$\cdot$10$^6$  had one hadron only.

The calorimeter is surrounded by an array of stations equipped with
scintillators in which the electromagnetic and muonic components of an air
shower are detected. A description of the experiment can be found in Antoni
\etal (2003).

\section{Simulations}

The measurements have been accompanied by extensive shower simulations in order
to understand the phenomenon of single hadrons and to determine the relation
between the primary proton spectrum and the single hadron spectrum at ground
level. The program CORSIKA~6.014 (Heck \etal 1998) has been employed with the
code QGSJET 01 (Kalmykov and Ostapchenko 1993, Kalmykov \etal 1997, Heck \etal
2001) for high-energy hadronic interactions and GHEISHA (Fesefeldt 1985) for
energies below 80~GeV.

\begin{figure}[t]
\epsfig{file=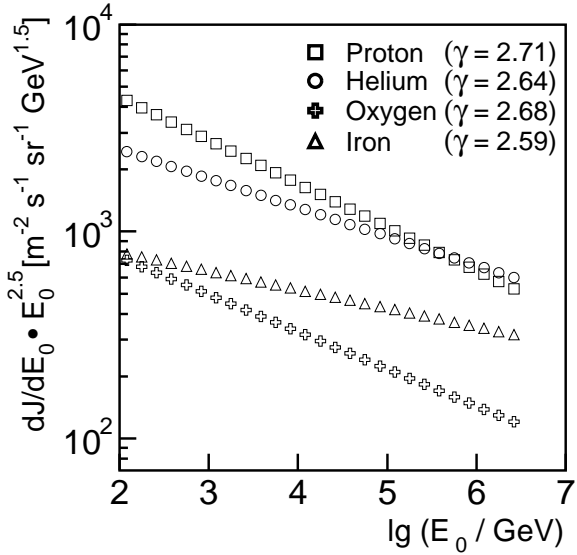,width=\columnwidth}
\caption{Primary flux values for the indicated nuclei vs. the primary particle
energy, according to parametrizations for oxygen (Wiebel-Sooth \etal 1998) and
for proton, helium, and iron (Hörandel 2003).  The corresponding spectral
indices $\gamma$ are given.}
\label{fig2}
\end{figure}

For the primary fluxes of nuclei above the atmosphere parametrizations
according to power laws are taken from compilations by Wiebel-Sooth \etal
(1998) and Hörandel (2003).  In the latter review more recent measurements are
taken into account and the parametrizations for proton, helium and iron have
been updated.  In total, about 2$\cdot$10$^{10}$ events have been simulated in
the energy range from 100~GeV to 3~PeV for proton, helium, oxygen, and iron
induced showers.  The number corresponds to a data taking period of
approximately 80 days for the calorimeter acceptance. For illustration, the
fluxes used are depicted in figure \ref{fig2}, marked with the corresponding
spectral index $\gamma$ and extrapolated into the PeV region.  The shower cores
of the simulated events have been distributed evenly over the calorimeter area
extended by 2~m at all four sides.  The hadrons are tightly concentrated near
the shower axis. The simulated distance distribution can be parametrized by an
exponential function that falls off to 1/e within 4.3 m, nearly independent of
energy.  Using this lateral distribution, calculations show that out of all
primary particles, which are reconstructed as single hadrons 75\% have been
simulated.  The simulations reveal, though, that for a given interval of hadron
energy, the fraction of missing events is nearly independent of primary energy.
Since the missing percentage does not depend strongly on energy it has no
significant effect on the deduced primary proton spectrum which will be
presented in section 5 below.  
%Hence, it needs not to be taken into account.

\begin{figure}[t]
\epsfig{file=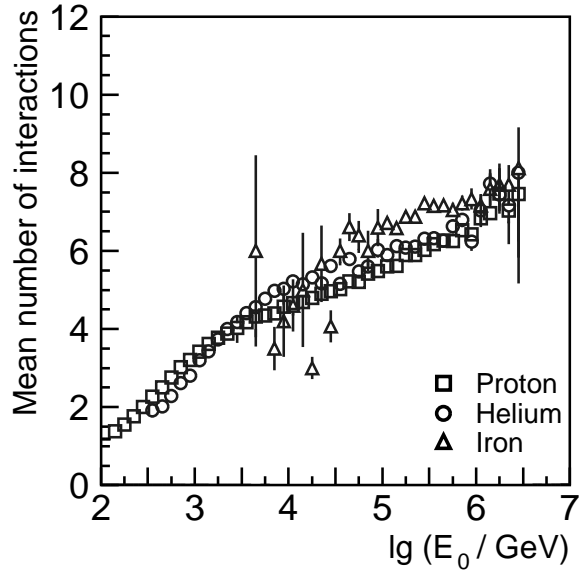,width=\columnwidth}
\caption{Mean number of hadronic interactions in single hadron events
vs. the primary energy for three primary particles as indicated.}
\label{fig3}
\end{figure}

\begin{figure*}\centering
\epsfig{file=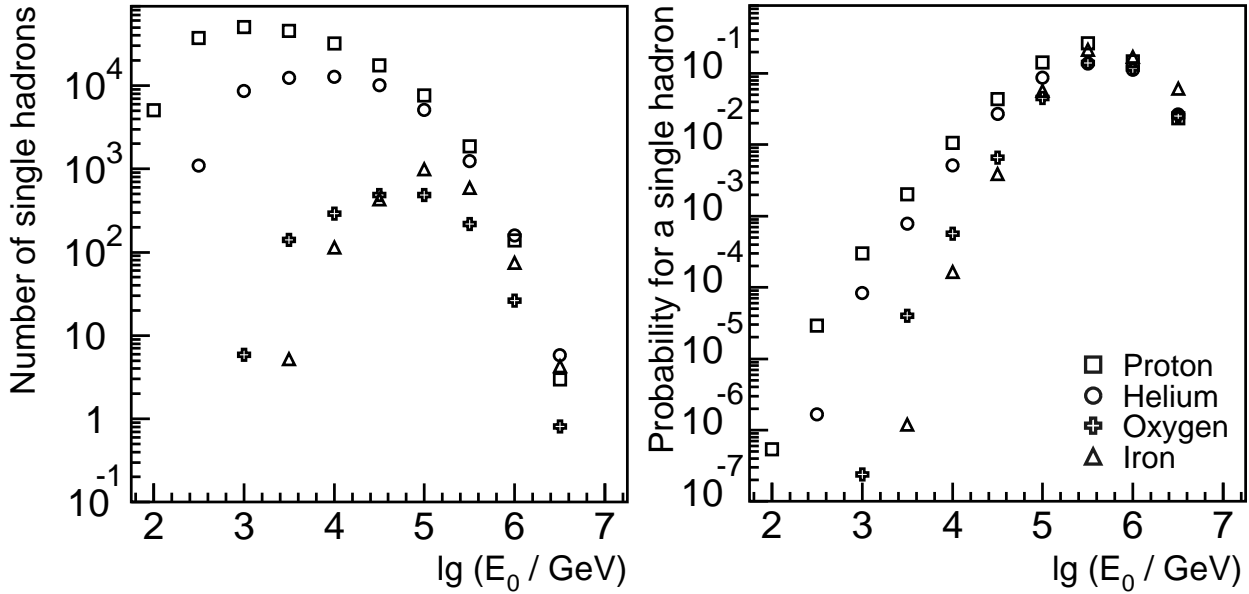,width=\textwidth}
\caption{Number of reconstructed single hadron events in the simulations vs.
primary energy for the indicated nuclei (left-hand). The probability to find a
single hadron vs.  primary energy is shown in the right-hand graph.}
\label{fig4}
\end{figure*}

Single hadrons belong to a particular class of air showers for which the
detected hadrons stem from air showers with only a few interactions in the
atmosphere. In vertical direction the thickness of the atmosphere corresponds
to about 11 interaction lengths. How many hadronic interactions occurred for
single hadron events is shown in figure \ref{fig3}. In the TeV range primary
protons or helium nuclei encounter 3 or 4 interactions only. Integrated over
all relevant energies, the average number of interactions is 3.6 for primary
protons.  Heavier nuclei do not generate single hadrons in this energy range,
as revealed by the primary iron simulations. Oxygen and iron nuclei can be seen
in this class of events only if the primaries have energies higher than 30~TeV.

The number of reconstructed single hadron events for the four classes of
primaries are shown in figure \ref{fig4} on the left-hand panel. On the
right-hand the corresponding probability to find a single hadron event is
plotted for primaries with respect to their energy.  One observes that up to a
few TeV, indeed, single hadrons originate mostly from primary protons and that
above 10~TeV also helium primaries contribute. At approximately 1~PeV, proton
and helium contribute with equal numbers to single hadrons and at higher
energies also heavier nuclei have to be considered. The right-hand panel
reveals that in the hundred TeV range 10\% of all showers are of the single
hadron type. This figure is valid for the present definition of a single
hadron. If also accompanying electromagnetic energy would exclude single
hadrons, they would be encountered less frequently, as outlined in the next
section.

\begin{figure*}\centering
\epsfig{file=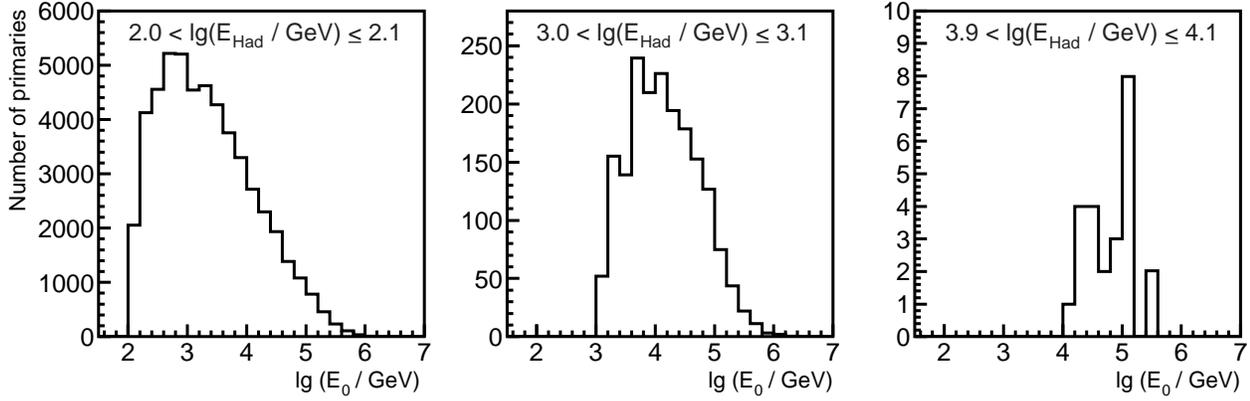,width=\textwidth}
\caption{Frequency distributions of primary energies for three intervals of
single hadron energies. Simulations with CORSIKA/QGSJET for proton initiated
showers.}
\label{fig5}
\end{figure*}

The simulations show how closely the single hadron events are related to the
primary particles. In figure \ref{fig5} the distributions of primary proton
energies are presented for three energy bins of single hadrons: for about
100~GeV, 1~TeV and 10~TeV.  The spread of the primary energy is wide, but the
bulk of parents have an energy ten times larger than the observed energy for
all three intervals.  Based on simulations it has been verified that the 25\%
of events missing do not change the shape of the distributions appreciably.
Therefore, despite the large fluctuations, measuring the single hadron spectrum
allows to deduce the primary proton flux.

\begin{figure*} \centering
\epsfig{file=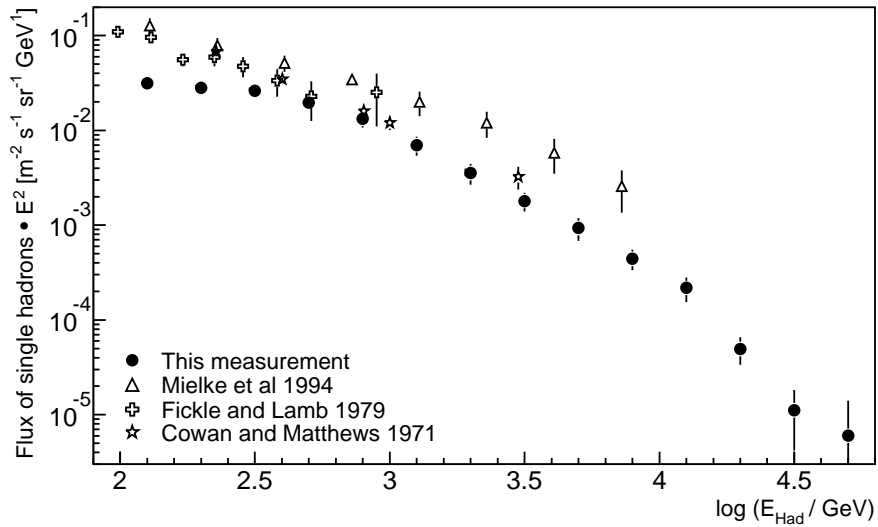,width=0.7\textwidth}
\caption{The single hadron spectrum: flux 
multiplied by the energy squared
vs. single hadron energy. For comparison 
results from other experiments near sea level are presented.}
\label{fig6}
\end{figure*}

\begin{table*}
\begin{center}
\begin{tabular}{ccr} \hline
lg (E$_{Had}$ / GeV) & ~ ~ number of hadrons & ~ ~ hadron flux (m$^2$ s sr GeV)$^{-1}$ \\
\hline
% 1.9 & 11527 & (~0.376~$\pm$~0.548~)~$\cdot~10^{-6}$ \\
2.1 & 834920 & (~0.20~$\pm$~0.03~)~$\cdot~10^{-5}~ ~ ~ ~$ \\
2.3 & 693430 & (~0.71~$\pm$~0.11~)~$\cdot~10^{-6}~ ~ ~ ~$ \\
2.5 & 481890 & (~0.26~$\pm$~0.04~)~$\cdot~10^{-6}~ ~ ~ ~$ \\
2.7 & 252110 & (~0.79~$\pm$~0.12~)~$\cdot~10^{-7}~ ~ ~ ~$ \\
2.9 & 113930 & (~0.21~$\pm$~0.04~)~$\cdot~10^{-7}~ ~ ~ ~$ \\
3.1 & 39510 & (~0.44~$\pm$~0.10~)~$\cdot~10^{-8}~ ~ ~ ~$ \\
3.3 & 13220 & (~0.89~$\pm$~0.22~)~$\cdot~10^{-9}~ ~ ~ ~$ \\
3.5 & 4400 & (~0.18~$\pm$~0.04~)~$\cdot~10^{-9}~ ~ ~ ~$ \\
3.7 & 1515 & (~0.37~$\pm$~0.10~)~$\cdot~10^{-10}~ ~ ~$ \\
3.9 & 450 & (~0.70~$\pm$~0.17~)~$\cdot~10^{-11}~ ~ ~$ \\
4.1 & 145 & (~0.14~$\pm$~0.04~)~$\cdot~10^{-11}~ ~ ~$ \\
4.3 & 23 & (~0.12~$\pm$~0.04~)~$\cdot~10^{-12}~ ~ ~$ \\
4.5 & 4 & (~0.11~$\pm$~0.07~)~$\cdot~10^{-13}~ ~ ~$ \\
4.7 & 2 & (~0.24~$\pm$~0.32~)~$\cdot~10^{-14}~ ~ ~$\\ \hline
\end{tabular}
\end{center}
\caption{Single hadron fluxes from the vertical direction
measured at sea level. The errors represent systematic uncertainties.}
\label{tab0}
\end{table*}

\section{Single hadron spectrum}
The flux of single hadrons is obtained using the trigger and reconstruction
efficiencies as determined with the GEANT/FLUKA code. The data are given in
table \ref{tab0} and are presented in figure \ref{fig6}.  The errors quoted are
estimated systematic uncertainties concerning the fiducial area of the
calorimeter, the total data taking time, the effective solid angle, and the
energy assignment, but are dominated by the trigger and reconstruction
efficiencies.  They amount to approximately 15\% below 1~TeV, 25\% in the TeV
range and 35\% above 10~TeV. The numbers of collected hadrons given in the
table indicate that statistical errors can be neglected below 10~TeV. The data
exhibit almost a power law in the double logarithmic graph.  However, on the
energy scale of over 2.5 orders of magnitude a gentle bend is apparent.  Such a
behaviour is observed in the simulations as well.  For reference, measurements
from the literature (Cowan and Matthews 1971, Fickle and Lamb 1979) are shown
as well as the data from the KASCADE prototype calorimeter (Mielke \etal 1994).
The last experiment exhibits somewhat larger fluxes compared with the present
data due to its smaller surface of 6~m$^2$. The KASCADE calorimeter with
304~m$^2$ fiducial area has a more efficient veto for multiple hadron
detection.  Especially for low hadron energies this may cause the differences.
The data of the two older experiments, which both had smaller apertures of
about 0.65 m$^2\cdot$ sr, show a similar shape.

\begin{figure*} \centering
\epsfig{file=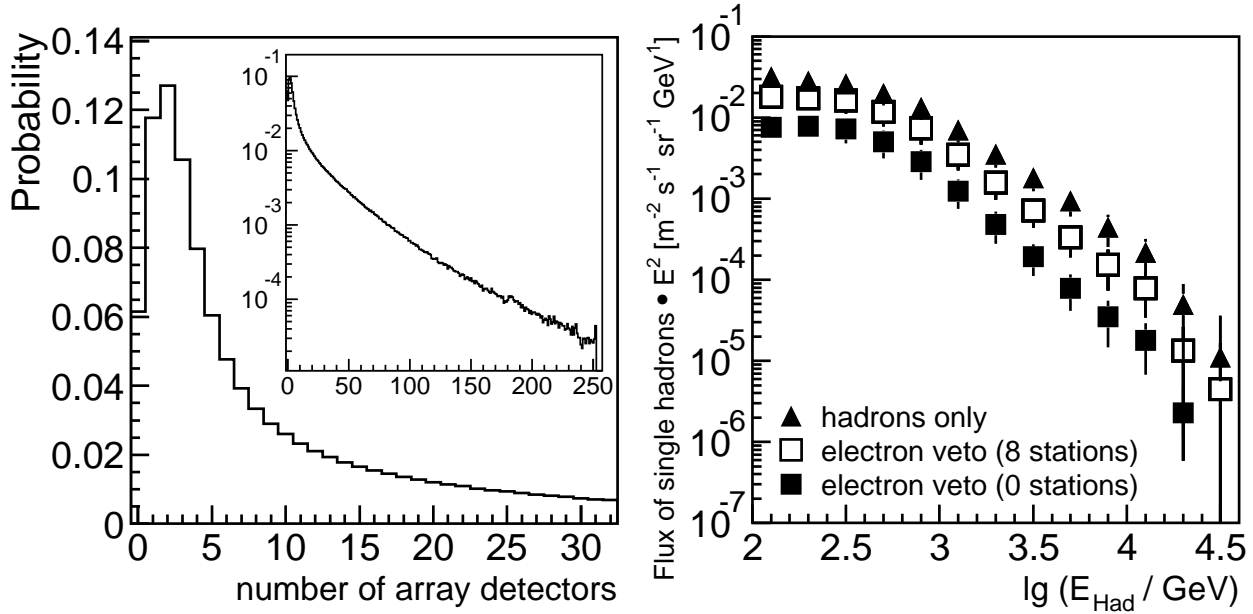,width=\textwidth}
\caption{Left-hand panel: Number of electromagnetic detectors with
E$_{dep}$~$\geq$~5~MeV.  Right-hand panel: The single hadron spectrum compared
with additional vetos: Not more than eight array stations or no station have
registered a minimum ionizing particle.  The ordinate has been multiplied by
the energy squared.}
\label{fig7}
\end{figure*}

In the present investigations also accompanying electromagnetic radiation
detected in the scintillators of the array stations can be accounted for.  The
graph in the left-hand panel of figure \ref{fig7} shows the number of stations
which have registered at least one minimum ionizing particle in coincidence
with a single hadron. In fact, only 6\% of the single hadron events are not
accompanied by a signal in the array stations.  As can be inferred from the
insert, a small probability exists that all 252 array stations have a signal in
coincidence with a single hadron. All this signifies that the notion ``single
hadron'' is somehow artificial, it depends on the experimental conditions and
ipso facto changes from experiment to experiment. Demanding that not more than
eight stations have an electronic signal, i.e. E$_{dep}$~$\geq$~5~MeV, already
reduces the number of single hadron events by 1/3. This can be seen in the
right-hand panel of figure \ref{fig7}, where the single hadron spectrum is
compared with this additional requirement.  One observes that the electron veto
becomes stronger for large hadron energies which in the mean originate from
higher primary energies.  Simulations indicate that the sensitivity to primary
protons is also enhanced.  In the figure the flux with no signal at all in the
252 stations is presented as well.  However, because the number of events
fulfilling this requirement is too small, the veto condition of not more than
eight stations activated has been chosen for further analyses, in particular
when deducing the primary proton spectrum.

\section{Primary proton spectrum}

\begin{figure}\centering
\epsfig{file=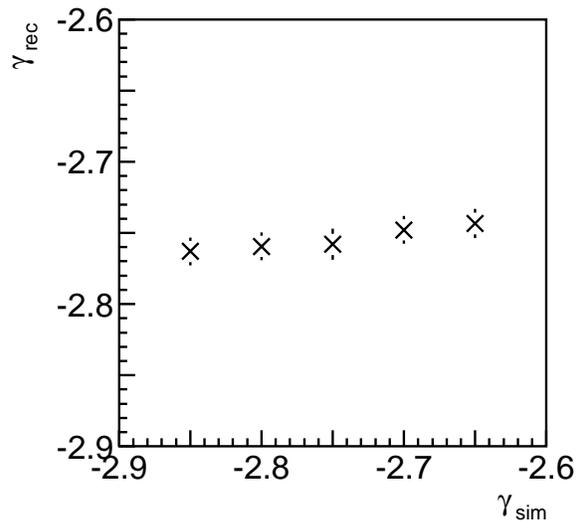,width=\columnwidth}                      
\caption{Spectral index of the primary proton spectrum deduced from the
measurements vs.  the slope used in the simulations of the probability density
distributions.
\label{fig8}}
\end{figure}

\begin{table*}
\begin{center}
\begin{tabular}{crr} \hline
primary energy E$_{0}$ & uncorrected flux $\cdot$ E$_{0}^{2.5}$ & 
proton flux $\cdot$ E$_{0}^{2.5}$ \\
lg(E$_{0}$/GeV) & 
~ ~ ~[m$^{-2}$s$^{-1}$sr$^{-1}$GeV$^{1.5}$] & 
~ ~ ~[m$^{-2}$s$^{-1}$sr$^{-1}$GeV$^{1.5}$] 
 \\
\hline
%2.0 & 5810~$\pm$~1430 & 5810~$\pm$~1430 \\
2.5 & 4680~$\pm$~1025~($\pm$~700) & 4675~$\pm$~1025~($\pm$~700) \\
3.0 & 3800~$\pm$~830~($\pm$~550) & 3400~$\pm$~740~($\pm$~500) \\
3.5 & 3000~$\pm$~660~($\pm$~450) & 2470~$\pm$~545~($\pm$~370) \\
4.0 & 2330~$\pm$~500~($\pm$~355) & 1800~$\pm$~395~($\pm$~270) \\
4.5 & 1900~$\pm$~565~($\pm$~490) & 1310~$\pm$~385~($\pm$~330) \\
5.0 & 1520~$\pm$~450~($\pm$~385) & 950~$\pm$~285~($\pm$~240) \\
5.5 & 1040~$\pm$~420~($\pm$~370) & 690~$\pm$~280~($\pm$~240) \\
6.0 & 775~$\pm$~310~($\pm$~285) & 505~$\pm$~200~($\pm$~180) \\ \hline
%6.5 & 870~$\pm$~560 & 390~$\pm$~250 \\ \hline
\end{tabular}
\end{center}
\caption{Primary flux of protons deduced from the single hadron spectrum assuming protons only
(second column) and with contributions from helium and heavy nuclei subtracted (third column).
The errors are systematic uncertainties (in parentheses the contribution from
the systematic errors of the single hadron spectrum).}
\label{tab1}
\end{table*}

The single hadron spectrum is converted to a flux of primary protons by
attributing to the measured hadrons a probable primary energy according to the
energy distributions which are illustrated in figure \ref{fig5}.  In principle
the single hadron spectrum $g(\lg {\rm E_{Had}})$ has to be converted to a flux
of primary protons $f(\lg {\rm E_0})$ by solving the integral equation
\begin{equation}
  g(\lg {\rm E_{Had}})=
  \int A(\lg {\rm E_{Had}}|\lg {\rm E_0})\ f(\lg {\rm E_0}) \
  d\lg \rm E_0.
\end{equation}
$A(\lg {\rm E_{Had}}|\lg {\rm E_0})$ is the transfer function transforming the
primary flux spectrum into the measured single hadron spectrum at ground level
accounting for the interactions in the atmosphere.  Several methods exist to
deconvolute one-dimensional spectra.  For sake of simplicity a slightly
different approach has been chosen, which turned out to be rather robust and
straightforward.  Knowing the probability density distribution $B(\lg {\rm
E_0}|\lg {\rm E_{Had}})$ for a given single hadron energy $\lg {\rm E_{Had}}$,
the primary proton flux can be inferred by 
\begin{equation}
  f(\lg {\rm E_0})=
  \int B(\lg {\rm E_0}|\lg {\rm E_{Had}})\ g(\lg {\rm E_{Had}}) \
  d\lg \rm E_{Had}.
\end{equation}

Using the calculated probability distributions (see e.g. figure \ref{fig5})
assuming primary fluxes as given in figure \ref{fig2}, the primary proton flux
is obtained. The resulting energy spectrum depends only slightly on the slope
of the a priori flux assumptions, as can be seen in figure~\ref{fig8}.  Plotted
are the power law indices derived from the data versus the assumed indices in
the simulations of the probability distributions.  One realizes that the method
yields rather stable results.

\begin{figure*}[t]\centering
\epsfig{file=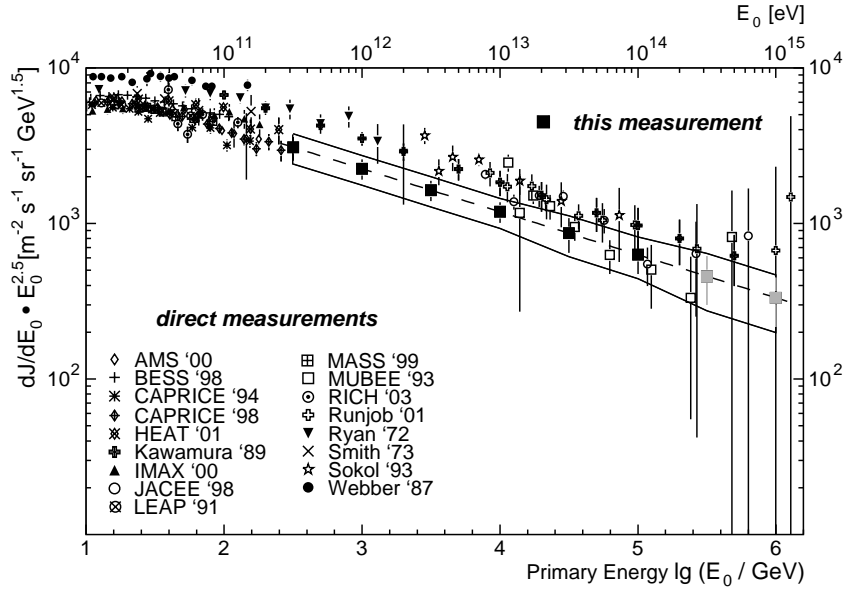,width=.7\textwidth}
\caption{The flux of primary protons as function of energy (black and grey
shaded squares). 
The dashed line represents a fit of a power law. The lines indicate the
maximum systematic errors estimated. For comparison results of direct
measurements are shown: AMS (Alcaraz \etal 2000), BESS (Sanuki \etal 2000), 
CAPRICE 94 (Boezio \etal 1999), CAPRICE 98 (Mocchiutti \etal 2001), 
HEAT (DuVernois \etal 2001), Kawamura (\etal 1989), IMAX (Menn \etal 2000),
JACEE (Asakimori \etal 1998), LEAP (Seo \etal 1991), 
MASS (Bellotti \etal 1999), MUBEE (Zatsepin \etal 1993),
RICH (Diehl \etal 2003), Runjob (Apanasenko \etal 2001), 
Ryan (\etal 1972), Smith (\etal 1973),
Sokol (Ivanenko \etal 1993), and Webber (\etal 1987).}
\label{fig9}
\end{figure*}

\begin{figure*} \centering
\epsfig{file=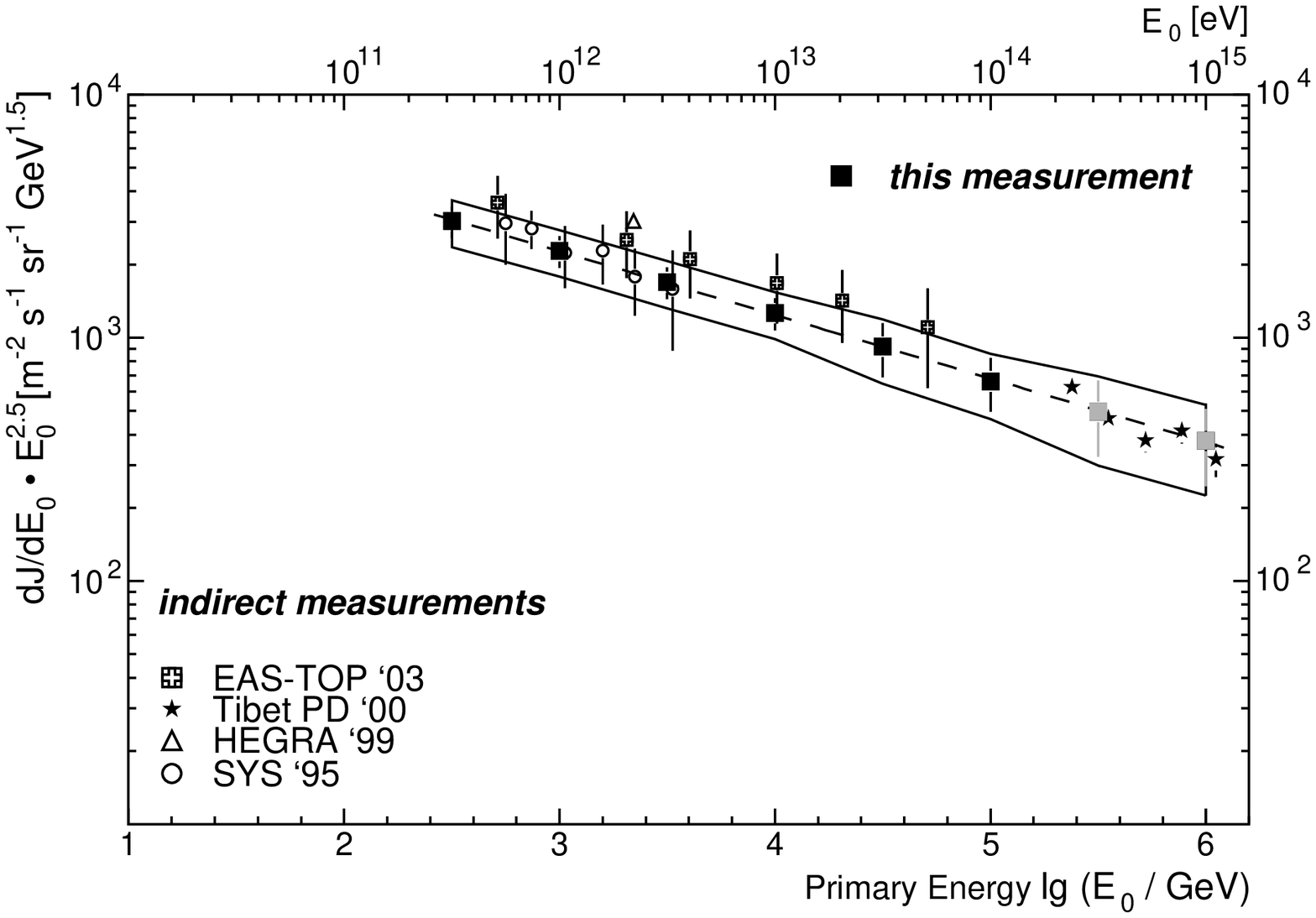,width=.7\textwidth}
\caption{Flux of primary protons as presented in figure \ref{fig9}.  For
comparison results of indirect measurements by air shower are shown: EAS-TOP
(Aglietta \etal 2003), Tibet (Amenomori \etal 2000), HEGRA (Aharonian \etal
1999), and SYS (Inoue \etal 1997).}
\label{fig10}
\end{figure*}

The outcome for the uncorrected primary fluxes are given in the second column
of table~\ref{tab1}, assuming that all primaries are protons. They are
multiplied by E$^{2.5}$ in order to compensate the steeply falling power law
spectrum. As seen in figure \ref{fig4}, at higher energies also helium and, to
some extent, the heavier nuclei initiate single hadron events. These
contributions are subtracted by taking into account the parametrized primary
fluxes as shown in figure \ref{fig2}.  The final proton fluxes are presented in
the third column. In the TeV region the difference of the flux values amounts
to about 20\%, a value comparable to the errors as discussed in the following.

The errors quoted are estimates of systematic uncertainties. They include the
calculation of the probability density distribution and the errors in the
single hadron spectrum (given in parentheses), both added quadratically.  The
uncertainties in $B(\lg {\rm E_0}|\lg {\rm E_{Had}})$ are estimated to amount
to 15\% below a few TeV and 20\% at higher values of E$_0$.

Graphic representations of the results are shown in figures \ref{fig9} and
\ref{fig10}. The present data are plotted as filled squares and the maximal
errors are indicated by the two lines.  The error bars on the individual points
represent the systematic uncertainties in the single hadron flux.  The squares
follow a power law reasonably well, a corresponding fit yields $dJ/dE_{0}~
=~(0.15\pm0.03)~\cdot~E_{0}^{-2.78\pm0.03}~$m$^{-2}$~s$^{-1}$~sr$^{-1}$~TeV$^{-1}$
which is indicated by the dashed line.  It should be kept in mind that for
higher energies above 100~TeV contributions of helium and heavy nuclei of up to
50\% had to be subtracted.  In the figures the corresponding values are marked
as shaded points.

In figure \ref{fig9} the data are compared to results of direct measurements
above the atmosphere taken from the literature.  One recognizes differences
between the individual results of the order of a factor of two.  Within these
uncertainties the present data are compatible with results from the literature.
At energies around 100~GeV the most recent data of direct measurements scatter
at the lower bound of the published fluxes. Our proton flux extrapolates well
to these data. This fact can be interpreted in such a way that in this energy
region from $10^2$ to $10^4$~GeV the hadronic shower cascade within the
atmosphere is well described by the program CORSIKA with the interaction code
QGSJET01. Also other tests have proven that below 1~PeV the latter code
describes the shower propagation best (Antoni \etal 1999 and 2001).

In figure \ref{fig10} our data are shown together with 
fluxes for primary protons obtained
by experiments using indirect methods of measurements as well.
Within the errors given, the data corroborate previous measurements.

\section{Summary}
Using the large hadron calorimeter of the KASCADE experiment during three years
of effective data taking, 2.5$\cdot$10$^6$ events have been accumulated for
which a single hadron was reconstructed. With these data the energy spectrum of
single hadrons has been derived.  These data are somewhat lower than fluxes
published previously. This is attributed to the large surface area of the
calorimeter which acts as a more efficient veto against multi-hadron events.

Single hadron events are particular air showers, which predominantly stem from
primary protons in the energy region considered. Applying large sets of
simulated single hadron events, and assuming primary particle fluxes as
obtained by direct measurements extrapolated into the PeV region by power laws,
the primary spectrum for protons could be determined. The data extend over
nearly four orders of magnitude in primary energy and can be described by a
power law with a flux parametrized as $dJ/dE_{0}~=~(0.15\pm0.03)~\cdot
E_{0}^{-2.78\pm0.03}~ $m$^{-2}$~s$^{-1}$~sr$^{-1}$~TeV$^{-1}$.
In the TeV region the proton fluxes agree well with recent measurements of
direct experiments above the atmosphere.

\section{Acknowledgments}
The authors would like to thank the members of the engineering and technical
staff of the KASCADE collaboration who contributed with enthusiasm and
commitment to the success of the experiment.  The KASCADE experiment is
supported by the German Federal Ministry of Education and Research and was
embedded in collaborative WTZ projects between Germany and Romania (RUM 97/014)
and Poland (POL 99/005) and Armenia (ARM 98/002).  The Polish group
acknowledges the support by KBN grant no. 5PO3B 13320.

\end{document}